\documentclass[twocolumn,showpacs,preprintnumbers,amsmath,amssymb]{revtex4}

\usepackage{graphicx}
\usepackage{dcolumn}
\usepackage{bm}

\begin{document}


\title{ Have Superkamiokande Really Measured the Direction of the Atmospheric Neutrinos which Produce Fully Contained Events and Partially Contained Events ?}

\author{E.Konishi $^{1}$, Y. Minorikawa$^{2}$, V.I.Galkin$^{3}$, \\
M.Ishiwata$^{4}$ and A.Misaki$^{5,6}$}

\affiliation
{\noindent $^1$ Department of Electronics and Information System Engineering, Hirosaki University, 036-8561, Hirosaki, Japan\\
$^2$ Department of Science, School of Science and Engineering, 
Kinki University, Higashi-Osaka, 577-8502 Japan\\
$^3$ Department of Physics, Moscow State University, 119992, Moscow, Russia\\
$^4$ Department of Physics, Saitama University, 338-8570, Saitama, Japan\\
$^5$ Advanced Research Institute for Science and Engineering, Waseda University, 169-0092, Tokyo, Japan\\
$^6$ Innovative Research Organization for the New Century, Saitama University, 338-8570, Saitama, Japan\\}

\date{\today}

\begin{abstract}
\small{{ Quasi Elastic Scattering (QEL) is the dominant source for producing both Fully Contained
 Events and Partially Contained Events in the Superkamiokande(SK) detector for the atmospheric
 neutrinos, in the range $\sim$0.1 GeV to $\sim$10 GeV. In the analysis of SK events, it is assumed that
 the direction of the incident neutrino is the same as that of the detected charged lepton.
   In the present letter, we derive the distribution function for the scattering angle of 
the charged leptons, their averaged scattering angle and their standard deviation due to QEL.
 Then, it is shown that the SK assumption for the scattering angle of the charged leptons
 in the QEL is not valid. Further, we examine the influence of the azimuthal angle of the charged
 leptons over their zenith angle.  As the result, we conclude that the zenith angle distribution
 of the neutrino under the SK assumption does not reflect the real zenith angle distribution of
 the atmospheric neutrino which produces Fully Contained Events and Partially Contained Events.
  This result has clear implication for attempts to detect neutrino oscillations from the analyses
 of Fully Contained Events and Partially Contained Events in Superkamiokande.

 }
}
\begin{center}

(Submitted to Physical Review D, Rapid Communication)
\end{center}

\end{abstract}


\maketitle

\section{Introduction}
In the experiment in which they try to detect the neutrino oscillation, by using the size of 
the Earth and measuring the zenith angle distribution of the atmospheric neutrino events,
 such as, Superkamiokande experiment[1] -- hereafter, simply SK --, it is demanded that
 the measurements of the direction of the incident neutrino are being carried out
 as reliably as possible.
  Among the experiments concerned on the neutrino oscillation, the analysis of Fully Contained 
Events in SK is regarded as mostly ambiguity-free one, because the essential information
 to extract clear conclusion is stored inside the detector. In SK, they assume that the direction
 of the neutrino concerned is the same as that of the produced charged lepton (hereafter,
 simply SK assumption)[2,3]. However, the SK assumption does not hold in the just energies concerned 
for neutrino events produced inside the detector, which is shown later. \\

    In the energy region where Fully Contained Events and Parially Contained Events (single ring
 events) are analysed, quasi elastic scattering of neutrino interaction(QEL) is the dominant source 
for the atmospheric neutrino concerned[4]
 
\section{The Differential cross section for QEL and the scattering angle of the charged lepton}    
  The differential cross section for QEL is given as follows [5]. \\

    \begin{eqnarray}
         \frac{{\rm d}\sigma}{{\rm d}Q^2} = 
         \frac{G_F^2{\rm cos}^2 \theta_C}{8\pi E_{\nu}^2}
         \Biggl\{ A(Q^2) \pm B(Q^2) \biggl[ \frac{s-u}{M^2} \biggr] \notag\\
                  +C(Q^2) \biggl[ \frac{s-u}{M^2} \biggr]^2 \Biggr\}.
    \end{eqnarray}

\noindent The signs + and - refer to $\nu_{\mu(e)}$ and $\bar{\nu}_{\mu(e)}$ for charged current(c.c.) 
interaction, respectively.     The $Q^2$ denotes four momentum transfer between the incident neutrino
 and the charged lepton. As for details of other symbols, see the text [5].
  The relation among $Q^2$ and $E_{\nu}$, the incident energy of 
  neutrino, $E_{\ell}$, the energy of the emitted charged lepton 
  ((anti)muon or (anti)electron) and $\theta_{\rm s}$, the scattering 
  angle of the charged lepton,  is given as

      \begin{equation}
         Q^2 = 2E_{\nu}E_{\ell}(1-{\rm cos}\theta_{\rm s}).
      \end{equation}

\noindent Also, the energy of the charged lepton is given by

      \begin{equation}
         E_{\ell} = E_{\nu} - \frac{Q^2}{2M}.
      \end{equation}

For a given energy $E_{\nu}$ of the incident neutrino, we randomly sample $Q^2$ through the Monte Carlo
 procedure from Eq. (1). Subsequently we obtain the scattering angle $\theta_{\rm s}$ of the charged lepton
 concerned by Eqs. (2) and (3). Thus, we obtain the distribution functions for scattering angle of
 the charged lepton. In Fig. 1, we give such distribution functions for different incident 
neutrino energies. Through such a Monte Carlo procedure, we obtain the average scattering angles
 and their standard deviations, too. We give them in Table 1. It is shown clearly from the figure
 and the table that the average scattering angles largely deviate from the direction of 
the incident neutrino, being accompanied by rather large standard deviations and consequenly
 we cannot neglect the scattering angle in the energy region where SK was interested in, 
say $\sim$0.1 GeV to $\sim$10 GeV.\\

\begin{figure}
\vspace{-4mm}
\includegraphics[scale=.3,angle=90]{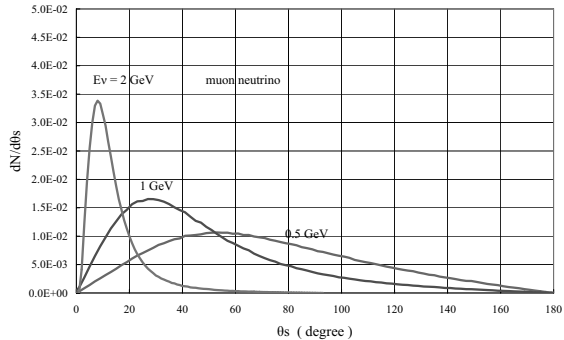}
\vspace{-12mm}
\caption{\label{fig:1} Distribution function for the scattering angle $\theta_{\rm s}$ of the muon for muon-neutrino. 
}
\end{figure}

\begin{center}
\begin{table}
\caption{\label{tab:table1} The average values $<\theta_{\rm s}>$ for scattering angle of the 
emitted charged leptons and their standard deviations $\sigma_{\rm s}$ for various primary 
neutrino energies $E_{\nu}$}
\begin{tabular}{cccccc}
\hline
\hline

$E_{\nu}$ (Gev)&angle&$\nu_{\mu}$&$\bar{\nu_{\mu}}$&$\nu_e$&$\bar{\nu_e}$ \\
&(degree)&&&&\\
\hline
0.2&$<\theta_\mathrm{s}>$&\ \ 89.86\ \ &\ \ 67.29\ \ &\ \ 89.74\ \ &\ \ 67.47\ \  \\
&$\sigma_\mathrm{s}$ & 38.63 &36.39 & 38.65&36.45 \\
0.5&$<\theta_\mathrm{s}>$&72.17&50.71&72.12&50.78 \\
&$\sigma_\mathrm{s}$ &37.08 &32.79 &37.08&32.82 \\
1&$<\theta_\mathrm{s}>$&48.44&36.00&48.42&36.01 \\
&$\sigma_\mathrm{s}$ &32.07&27.05 & 32.06&27.05 \\
2&$<\theta_\mathrm{s}>$&25.84&20.20&25.84&20.20 \\
&$\sigma_\mathrm{s}$ & 21.40 &17.04 & 21.40&17.04 \\
5&$<\theta_\mathrm{s}>$&8.84&7.87&8.84&7.87 \\
&$\sigma_\mathrm{s}$ & 8.01 &7.33 & 8.01&7.33 \\
10&$<\theta_\mathrm{s}>$& 4.14&3.82&4.14&3.82 \\
&$\sigma_\mathrm{s}$ & 3.71 &3.22 & 3.71&3.22 \\
100&$<\theta_\mathrm{s}>$&0.38&0.39&0.38&0.39 \\
&$\sigma_\mathrm{s}$ & 0.23 &0.24 & 0.23&0.24 \\
\hline
\hline
\end{tabular}
\end{table}
\end{center}

\section{The influence of the azimuthal angle of the charged lepton over their zenith angle}
In addition to the scattering angle of the charged leptons, it should be
 emphaized that the azimuthal
 angles of the charged particles in QEL play a decisive role in the determination of their zenith
 angles as well as the translation from Fully Contained Events to Partially Contained Events 
(vice versa) which are mentioned later.\\ 

In order to examine the influence of the azimuthal angle of the charged leptons over their zenith
 angle, let us denote the direction cosines of the incident neutrino $( \ell,m,n )$ and denote the 
scattering angle of the charged lepton,  $\theta_{\rm s}$, and the azimuthal angle, $\phi$, with regard to the axis of
 the incident neutrino.
 Then,
  $( \ell_{\rm r}, m_{\rm r}, n_{\rm r} )$ , the direction cosines of the charged lepton 
which correspond to $( \ell,m,n )$ are given as
  
\begin{equation}
\left(
         \begin{array}{c}
             \ell_{\rm r} \\
             m_{\rm r} \\
             n_{\rm r}
         \end{array}
       \right)
           =
       \left(
         \begin{array}{ccc}
            \displaystyle \frac{\ell n}{\sqrt{\ell^2+m^2}} & -\displaystyle \frac{m}{\sqrt{\ell^2+m^2}} & \ell \\
            \displaystyle \frac{mn}{\sqrt{\ell^2+m^2}} & \displaystyle \frac{\ell}{\sqrt{\ell^2+m^2}} & m \\
                        -\sqrt{\ell^2+m^2} & 0 & n
         \end{array}
       \right)
       \left(
          \begin{array}{c}
            {\rm sin}\theta_{\rm s}{\rm cos}\phi \\
            {\rm sin}\theta_{\rm s}{\rm sin}\phi \\
            {\rm cos}\theta_{\rm s}
          \end{array}
       \right),
\end{equation}

\noindent while SK assume

    \begin{equation}
(\ell_r,m_r,n_r)=(\ell,m,n)
    \end{equation}

\section{ Monte Carlo procedure, taking account of the effect of the azimuthal angle and discussions} 

  By using Eq. (4), we carry out a Monte Carlo calculation to examine the influence of the azimuthal
 angle of the charged leptons over their zenith angle.
The scatter plots between $\cos{\theta}$, cosines of the zenith angles of the charged
 leptons and fractional energies $E_{\mu}/E_{\nu}$ of the charged leptons for diffrent directions of the incident neutrinos 
 are given in Figs. 2 to 4. For a given $Q^2$ in Eq. (1), the energy $E_{\ell}$ of the charged lepton and its scattering angle $\theta_{\rm s}$ is uniquely determined due to the two body kinematics. In Fig. 2, we give
 the case of vertically incident neutrinos $(\theta_{\nu}=0^{\rm o})$. Here, as the zenith angles of the charged leptons are
 measured from the vertical direction in the SK case, the azimuthal angles of the charged leptons
 never influence over their zenith angle, and consequently the relation 
 between their fractional
 energies and their zenith angles is uniquley determined as mentioned above. In Fig. 3, we give the case of horizontally incident neutrinos $(\theta_{\nu}=90^{\rm o})$.
  Here, the azimuthal angle of the charged leptons has a potent influence on their zenith angle through the operation of Eq. (4).
   As is seen clearly from the figure, the $\cos{\theta}$ is widely distributed even to the backward for the same energy of the charged lepton. In Fig. 4, we give the intermediate case of the diagonal incidence $(\theta_{\nu}=43^{\rm o})$.  \\

\begin{figure}
\vspace{-6mm}
\includegraphics[scale=.3,angle=-90]{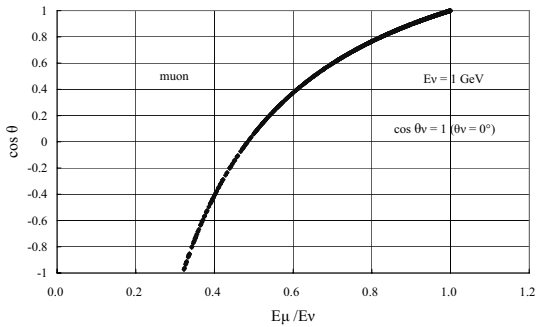}
\vspace{-12mm}
\caption{\label{fig:2} The scatter plot between the fractional energies  $E_{\mu}/E_{\nu}$ of the emitted muons and 
the cosine of zenith angles, ${\rm cos}\theta$, for muon neutrinos with 1 GeV. The incident direction of the neutrinos is vertical. 
The sampling number is 1000.
}
\end{figure}

\begin{figure}
\vspace{-6mm}
\includegraphics[scale=.3,angle=-90]{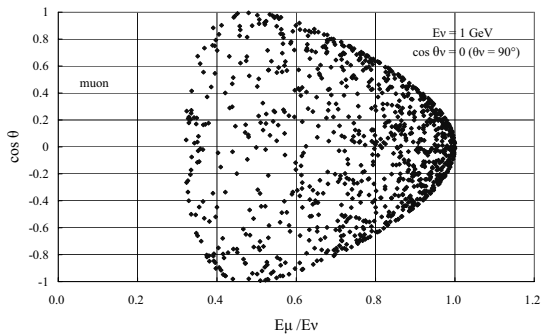}
\vspace{-12mm}
\caption{\label{fig:3} The scatter plot between $E_{\mu}/E_{\nu}$ and ${\rm cos}\theta$. The incident direction of the neutrinos is horizontal. The other quantities are the same as in Fig. 2.}
\end{figure}

\begin{figure}
\vspace{-6mm}
\includegraphics[scale=.3,angle=-90]{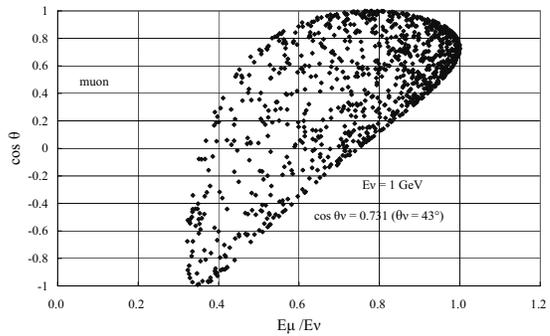}
\vspace{-12mm}
\caption{\label{fig:4} The scatter plot between $E_{\mu}/E_{\nu}$ and ${\rm cos}\theta$. The incident direction of the neutrinos is diagonal. The other quantities are the same as in Fig. 2.
The incident direction of the neutrinos is diagonal. The other quantities are 
the same as in Fig. 2.
}
\end{figure}

To connect our results with the analysis of the real experimental data , we finally need to take account of  
the energy spectrum of the incident neutrino in our calculation. For this purpose, we adopt
 the neutrino energy spectrum at Kamioka site obtained by Fiorentini {\it et al.}[6] and have carried out
 the following Monte Carlo procedure for a given ${\cos \theta_{\nu}}$ of the incident neutrino .\\
  
\noindent Procedure A: we randomly sample the energy of the incident neutrino from the probability 
function which is composed of the combination of the neutrino energy spectrum by Fiorentini {it et al.}, which covers from 0.1 GeV to 100 GeV at Kamioka site , with the corresponding total cross section for QEL.\par
\noindent Procedure B: we decide whether the neutrino concerned should be attributed to the particle
 or the anti-particle by the random sampling from the corresponding probability functions given in Procedure A.\par
\noindent Procedure C: we randomly sample $Q^2$ of the neutrino concerned for a given energy of the 
neutrino from Eq. (1).\par

\noindent Procedure D: we decide the energy of the charged lepton, $E_{\ell}$, from Eq. (3) and
 its scattering angle,${\theta_s}$, from Eq. (2).\par   

\noindent Procedure E: we randomly sample the azimuthal angle, ${\phi}$, by using a random number
 between (0,1).\par

\noindent Procedure F:  we decide the direction cosines of the charged leptons from Eq. (4), being 
accompanied by the scattering angle and the azimuthal angle obtained from Procedures D and E.\\

We repeat a chain of the procedures A to F until we attain at the required trial number
 (ten thousands samplings per each case).\\

\begin{figure}
\vspace{-6mm}
\includegraphics[scale=.3,angle=90]{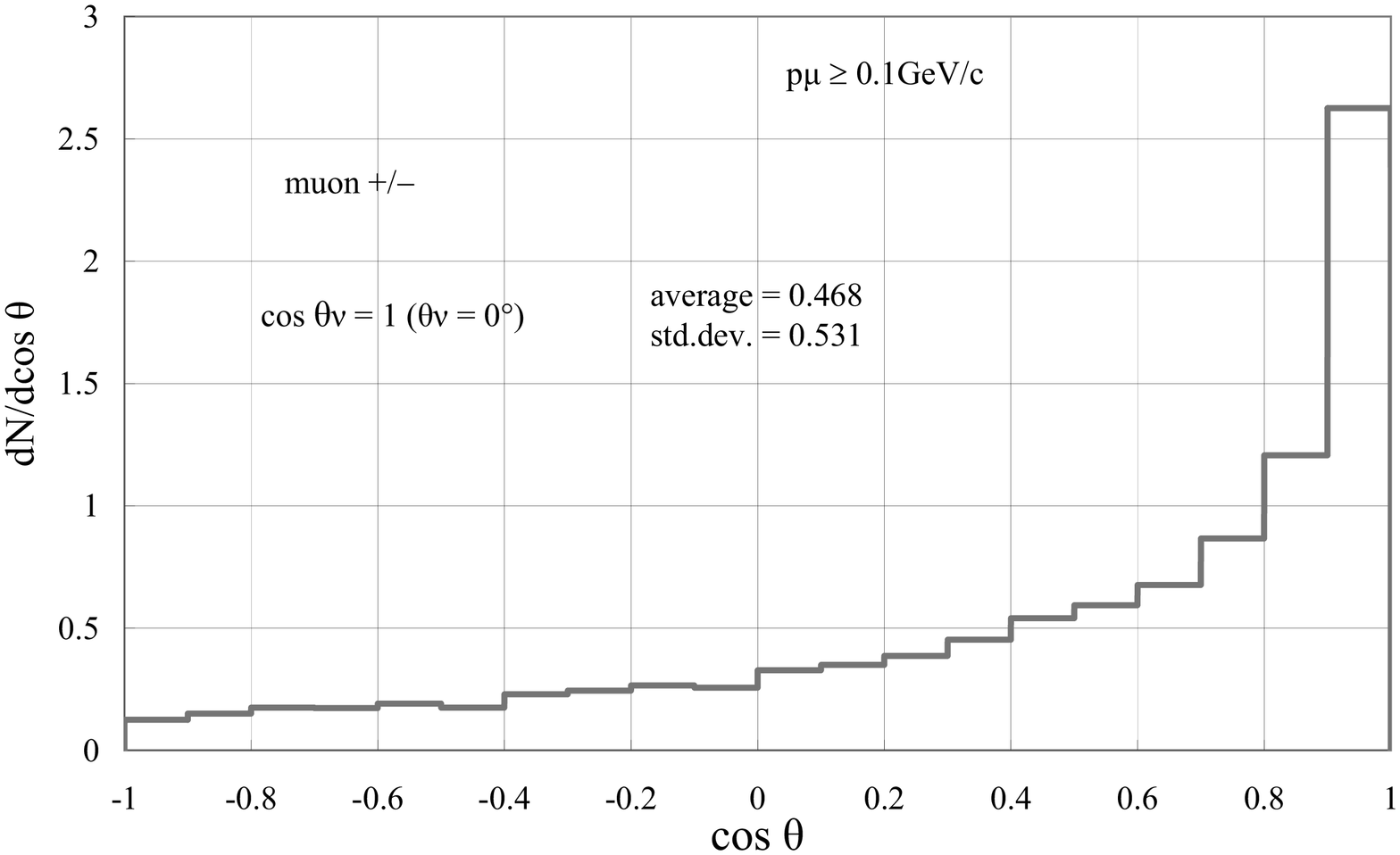}
\vspace{-12mm}
\caption{\label{fig:5} Zenith angle distribution of $\mu^+$ and $\mu^-$ for $\nu_\mu$ and $\nu_\mu$, 
taking account of the overall neutrino spectrum at Kamioka site. The direction of the incident neutrino
 is vertical.
}
%
%
\vspace{-3mm}
\includegraphics[scale=.3,angle=90]{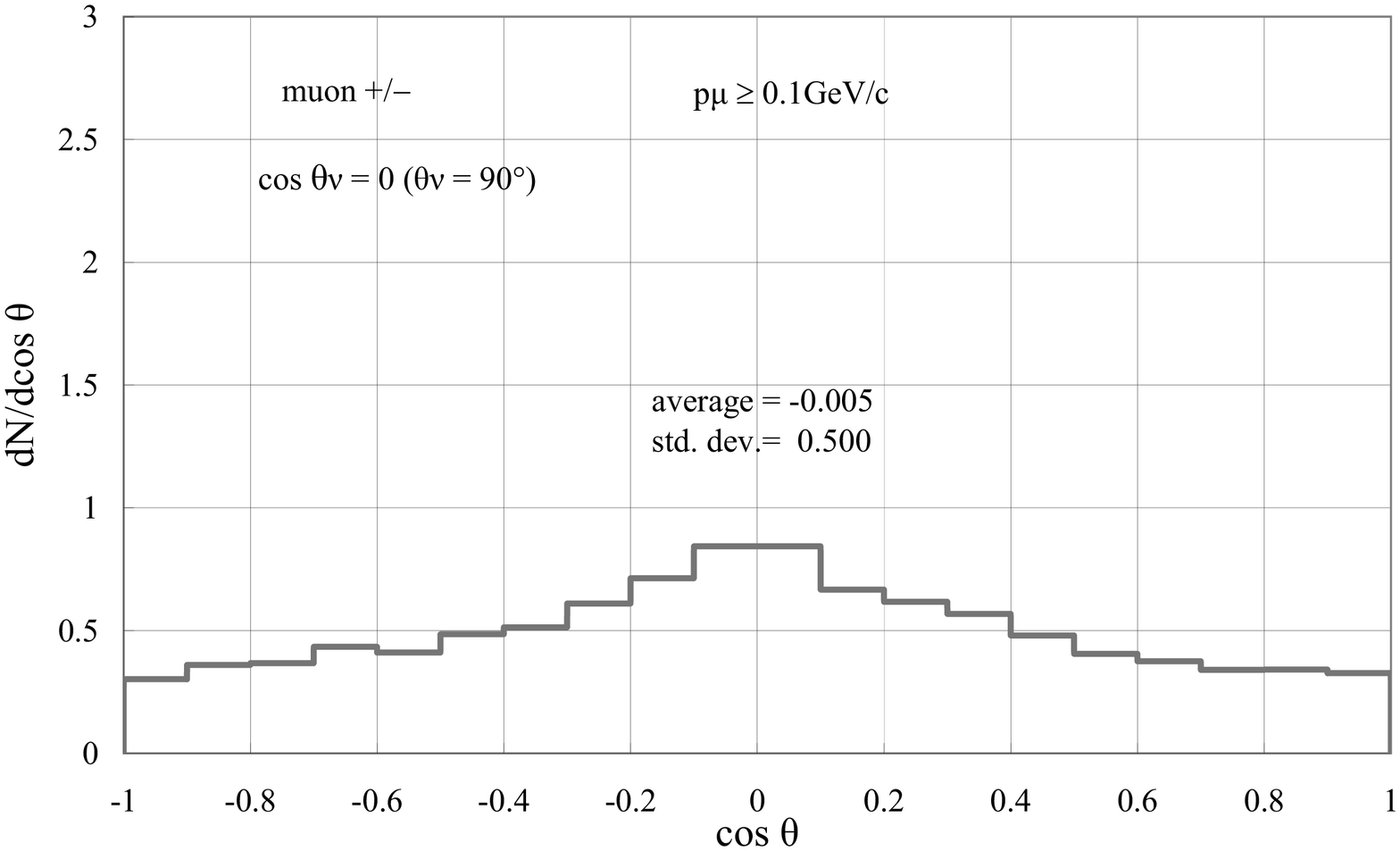}
\vspace{-12mm}
\caption{\label{fig:6} Zenith angle distribution of $\mu^+$ and $\mu^-$ for $\nu_\mu$ and $\nu_\mu$, 
taking account of the overall neutrino spectrum at Kamioka site.The direction of the incident neutrino is horizontal. The other quantities are 
the same as in Fig. 5.
}
%
\vspace{-6mm}
\includegraphics[scale=.3,angle=90]{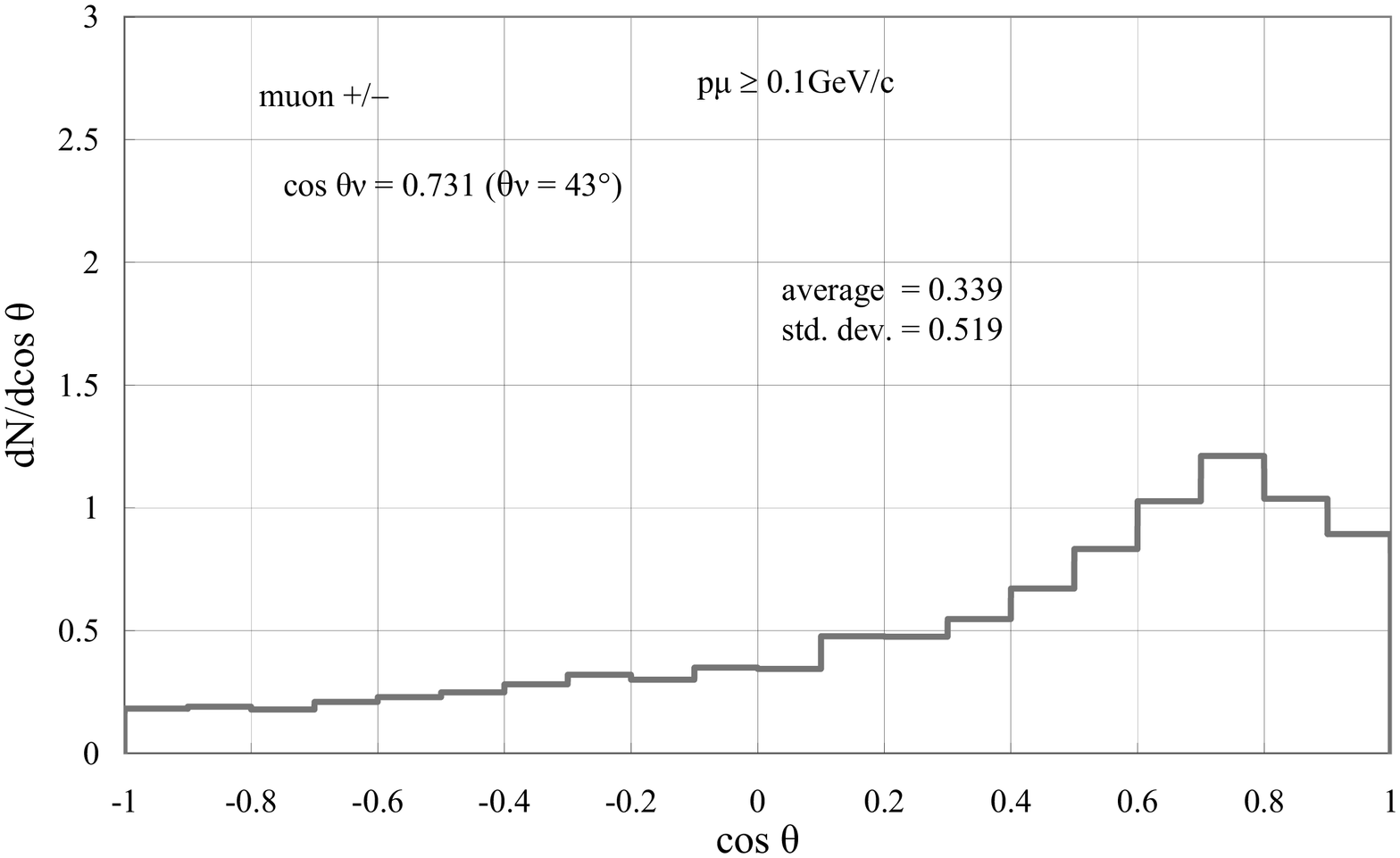}
\vspace{-12mm}
\caption{\label{fig:7} Zenith angle distribution of $\mu^+$ and $\mu^-$ for $\nu_\mu$ and $\nu_\mu$, 
taking account of the overall neutrino spectrum at Kamioka site.The direction of the incident neutrinos is diagonal. The other quantities are 
the same as in Fig. 5.
}
\end{figure}

  In Figs. 5 to 7 for the three cases of $\theta_{\nu}$ ($\theta_{\nu}=0^{\rm o},90^{\rm o}$ and $43^{\rm o}$), we give the zenith angle distribution of the sum of ${\mu^+}$ 
  and ${\mu^-}$ for a given  ${\theta_{\nu}}$, obtained by Procedures A to F. 
If the SK assumption is valid, even if it may be of approximation, the zenith angle distribution
 of the charged lepton should be of the delta function type with a peak around the direction of
 the incident neutrino. However, it is clear from the figures that the real zenith angle
 distributions of the charged leptons deviate far from the delta function type distribution, and
 further the backward scattering is too serious to neglect, which means rather large mixture
 of down going muon events into upward going events. It is also noticed that
 the azimuthal angles critically influence the translation from Fully Contained Events to 
Partially Contained Events (vicer versa)  which is strongly dependent on their generation point
 inside the detector.\
\begin{figure}
\vspace{-6mm}
\includegraphics[scale=.3,angle=90]{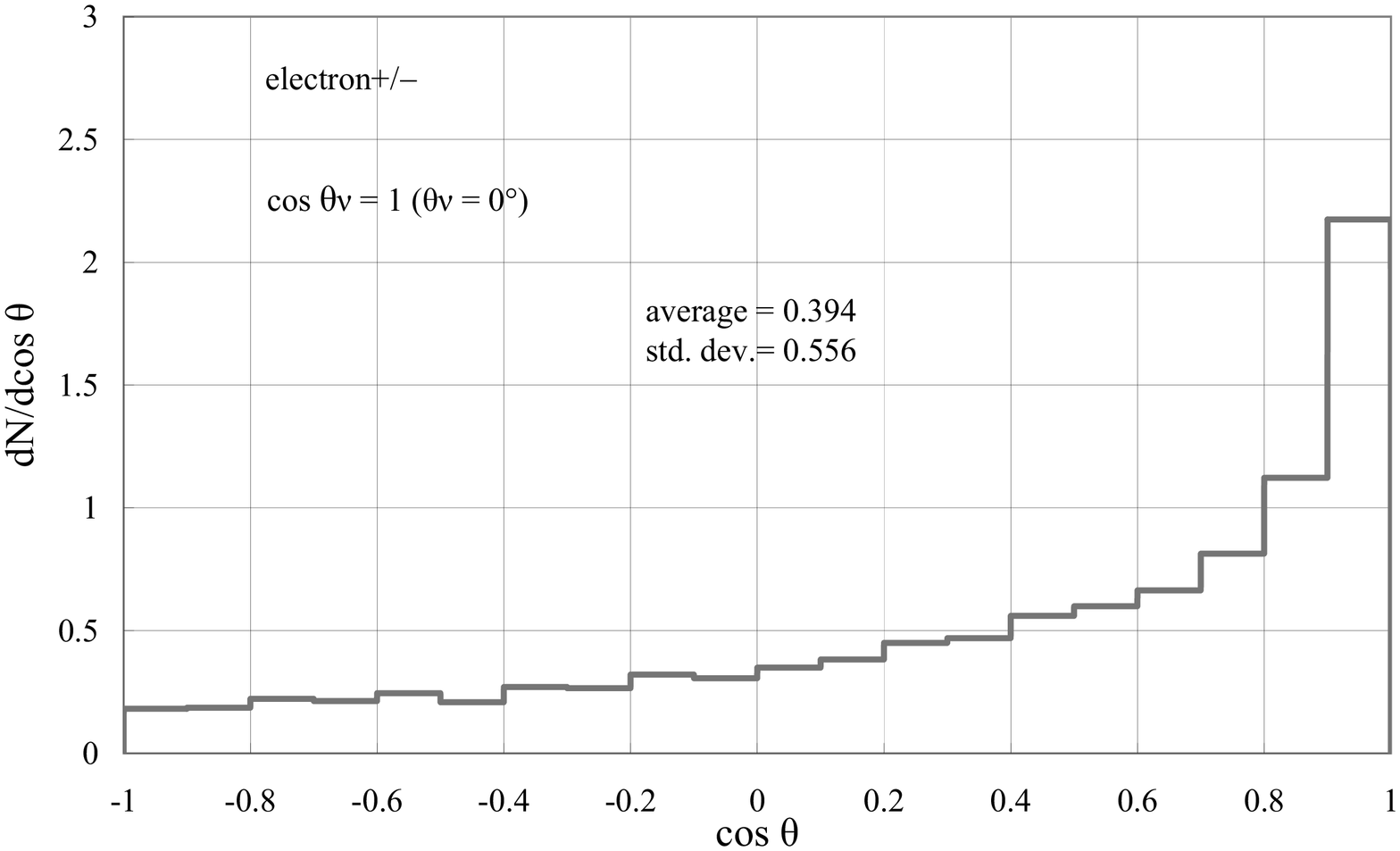}
\vspace{-12mm}
\caption{\label{fig:8} Zenith angle distribution of $e^-$ and $e^+$ for $\nu_e$ and 
$\bar{\nu_e}$, taking account of the overall neutrino spectrum at Kamioka site.The direction 
of the incident neutrino is vertical. The sampling number is 10000.
}
\vspace{-6mm}
\includegraphics[scale=.3,angle=90]{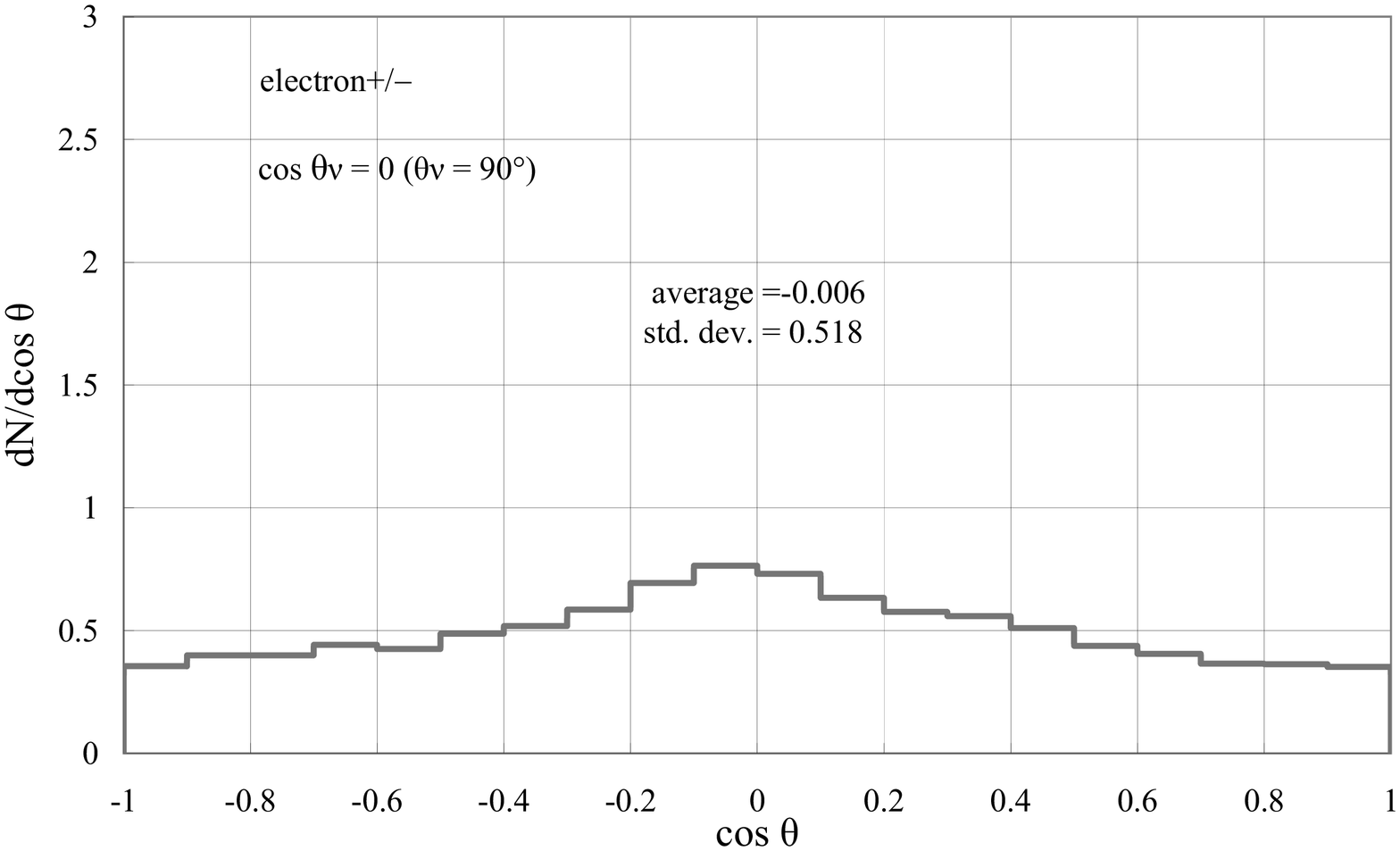}
\vspace{-12mm}
\caption{\label{fig:9} Zenith angle distribution of $e^-$ and $e^+$ for $\nu_e$ and 
$\bar{\nu_e}$, taking account of the overall neutrino spectrum at Kamioka site.
 The direction of the incident neutrino is horizontal. The other 
quantities are thesame as in Fig. 8.
}
\vspace{-6mm}
\includegraphics[scale=.3,angle=90]{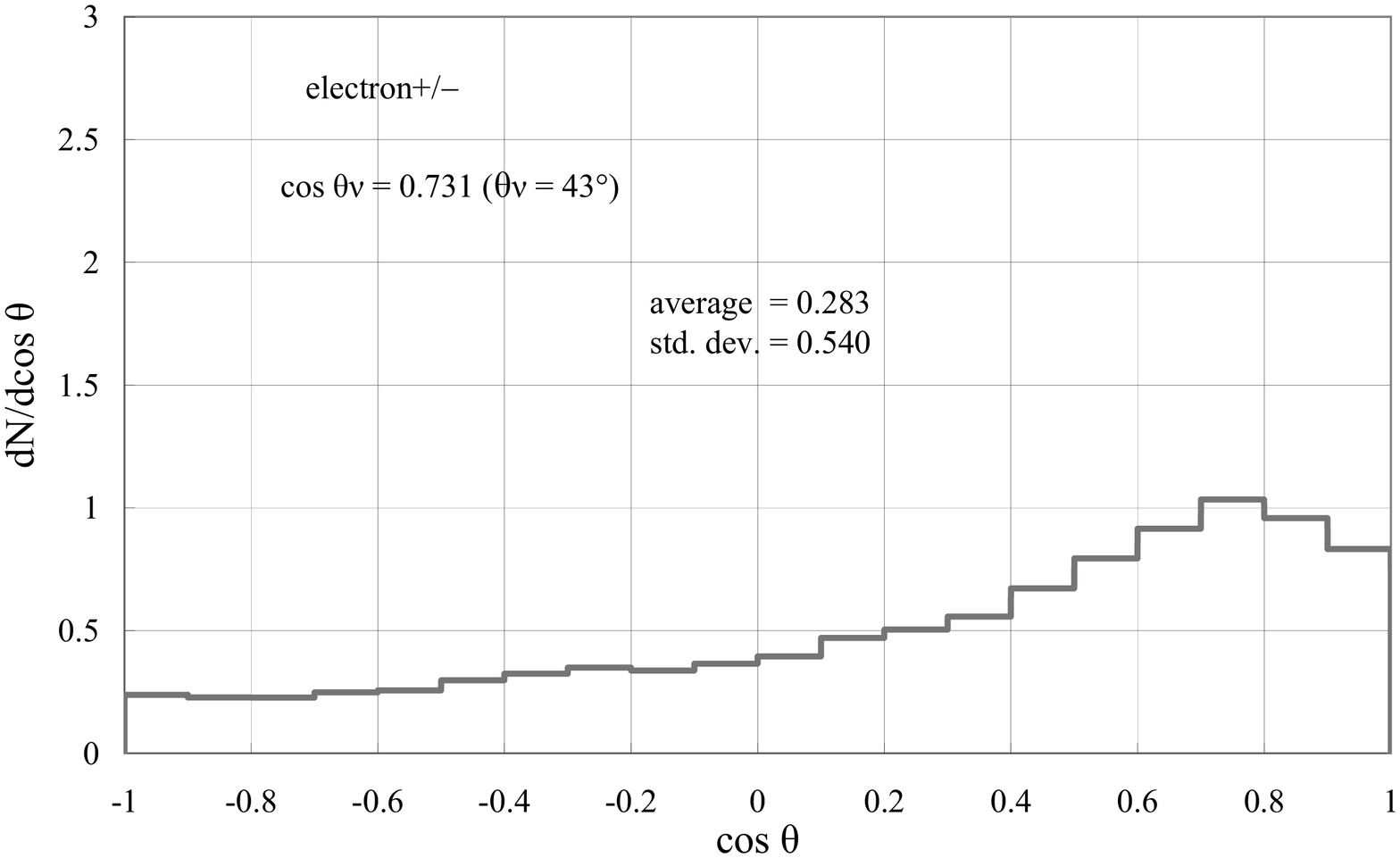}
\vspace{-12mm}
\caption{\label{fig:10}  Zenith angle distribution of $e^-$ and $e^+$ for $\nu_e$ and 
$\bar{\nu_e}$, taking account of the overall neutrino spectrum at Kamioka site.
The direction of the incident neutrino is diagonal. The other 
quantities are the same as in Fig. 8.
}
\end{figure}

In Figs. 8 to 10, we give similar results for $e^-$ and $e^+$ which correspond to Figs. 5 to 7.

The charcteristics of the Figs. 8 to 10 for (anti-)electron is essentially the same as in 
Figs. 5 to 7 for (anti-)muon.

\section{conclusion}   
   From these figures 5 to 10, it is surely concluded that the zenith angle
    distributions of the charged leptons
 under the SK assumption does not reflect the real zenith angle distribution of
  the incident neutrinos concerned. Of course, it is common sense that the
   direction of individual charged lepton cannot determine that of the
    neutrino concerned uniquely, but only statistically. 
      However, our present results clearly shows that the direction of the charged 
 leptons under the SK assumption cannot dtermine those of the incident neutrino, even if they 
 accumulate experimental data statistically enough. Thus, it is surely concluded that we 
 could not extract any definite conclusion around the neutrino oscillation from the 
 measurement on the charged leptons under SK assmption  which occur inside the detector and is regarded as the most reliable one from the point of the experimental accuracy. 

   The relation between the zenith angle distributions of the charged leptons which are really measured experimentally and those  of the parent neutrinos, taking account of the scattering angle and the azimuthal angle of the charged leptons correctly, will be shown in a subsequent paper.


\end{document}